# Enhanced Pauli spin response, failure of Stoner & spin fluctuation models, and presence of 6 $eV$ plasmonic excitations in Ni metal

Shivani Bhardwaj[1,*] Antik Sihi[1,†] and Sudhir K. Pandey[2,‡]
[1]*School of Physical Sciences, Indian Institute of Technology Mandi, Kamand - 175075, India and*
[2]*School of Mechanical and Materials Engineering,*
*Indian Institute of Technology Mandi, Kamand - 175075, India*

We revisit the electronic structure of Nickel (Ni) using the density functional theory (DFT) and dynamical mean-field theory (DMFT) for the theoretical description of its electronic structure properties along with finite-temperature magnetism. Our study provides a comprehensive account of electronic and magnetic properties with the same set of Coulomb interaction parameters, $U$=5.78 $eV$ and $J$=1.1 $eV$ calculated using first-principles approach. The nature of theoretical magnetization curves obtained from DFT and DFT+DMFT as well as the experimental curve show deviation from the standard models of magnetism, *viz* Stoner and spin fluctuation model. In comparison to DFT+DMFT method, temperature dependent DFT approach is found to well describe the finite-temperature magnentization curve of Ni below critical temperature ($T \leq 631$ K). The study finds significant Pauli-spin susceptibility contribution to paramagnetic spin susceptibility. Excluding the Pauli-spin response yields a linear Curie-Weiss dependence of the inverse paramagnetic susceptibility at higher temperatures. Also, the presence of mixed valence electronic configuration ($3d^8$, $3d^9$ and $3d^7$) is noted. The competing degrees of both the itinerant and localized moment picture of $3d$ states are found to dictate the finite-temperature magnetization of the system. Furthermore, the quasiparticle scattering rate is found to exhibit strong deviation from $T^2$ behavior in temperature leading to the breakdown of conventional Fermi-liquid theory. In addition to the 6 $eV$ feature, our calculated electronic excitation spectrum confirms the satellite feature extending ∼10 $eV$ binding energy, being consistent with experimental observation. Interestingly, our $G_0W_0$ results find the presence of plasmonic excitation contribution to the intensity of famous 6 $eV$ satellite along with the electronic correlation effects, paving way for its reinterpretation.

## I. INTRODUCTION

The theoretical understanding of metallic ferromagnets especially $3d$ transition metals has been a long standing problem for decades. The main difficulty in their theoretical description stems from the apparent dual character of their $3d$-electrons: they can be characterized as itinerant electrons by the band theory in their ground state, while their finite temperature properties strongly indicate the presence of local magnetic moments as evident from the exhibition of Curie-Weiss (CW) behavior of magnetic susceptibility[1,2].
Thereafter, theoretical studies have been focused on explaining the dual characteristics of $3d$ electron systems, considering the electron-electron correlation effects in the itinerant electron models[3,4]. Since several low temperature properties of $3d$ transition metals have been well illustrated by the band theory approximations such as Density functional theory (DFT) within LDA (local density approximation) or GGA (generalized gradient approximation), their electronic excitation spectra (EES) and finite temperature magnetic properties were long back recognized to pose challenge for their explanation at the DFT level[5–7].

The theoretical advancements in this direction saw the development of several beyond DFT methods i.e. $GW$, DFT+DMFT, to explain the spectral and magnetic properties of these $3d$ electron systems via the inclusion of electronic correlations in the $3d$ orbitals[8–14]. Evidently the most successful and accurate approach has been the use of DMFT in the framework of DFT+DMFT which adds the many-body effects as an effective quantum impurity problem.

Ni, in particular has been one of the extensively studied elemental transition metals pertaining to its non-trivial features of spectral plot *i.e.* the famous 6 $eV$ satellite feature, 30% narrowing of the occupied $3d$ bandwidth and 50% reduction of exchange splitting in comparison to DFT results and lower magnetic transition temperature ($T_c$) than that predicted by band theory[15,16]. Earliest of the reports on Ni realized the importance of electronic correlation effects ignored at the level of the band theories based schemes to account for its experimental observations[17,18]. For the past two decades numerous studies have been carried out on Ni, using DFT+DMFT approach with a variety of formulations based on- for instance, different types of impurity solvers ( eg. QMC (Quantum Monte Carlo)), Coulomb interaction parameterization (for eg. density-density etc.), self-consistency schemes typically considering empirically chosen set of Coulomb interaction parameters $U$ and $J$ (where $U$ stands for on-site Coulomb repulsion interaction, $J$ stands for Hund's like exchange interaction)[16,19–24]. Majority of the studies have been able to explain its broad aforementioned EES features. However, the finite temperature magnetism could only be explained in reduced temperature picture while lacking the quantitative account of finite temperature magnetization[16,24].
In order to account for the observed finite-temperature magnetic properties of $3d$ transition metals, the correctness of theoretical approach demands not only the explanation of observed $T_c$ but also the simultaneous explanation of $M(T)$. The theoretical studies concerning finite temperature magnetization properties are generally focused on predicting the $T_c$ apart from the saturation magnetization. The experimental $T_c$ of transition metals was found intractable by the band theory or the so called Stoner theory based on mean field and Hartree-Fock approximations and the explanation of $M(T)$ largely remained unsettled. The failure was recognized to be due to the lack of local moment formation picture in the band theory approach which could be addressed by including the spin-fluctuations of $3d$ electrons along with the itinerancy. Thus, the DFT+DMFT technique which includes the finite temperature spin fluctuations together with the zero temperature quantum spin fluctuations[25] was widely used approach for studying the finite-temperature magnetic properties of transition metals.

The significant overestimation of $T_c$ in the earliest of DFT+DMFT studies was attributed to the use of density-density type of paramterization of Coulomb interaction[16,26]. Nevertheless, the use of rotationally invariant type (Full type) of Coulomb interaction parametrization showed improved agreement with the observed $T_c$ but was found to be accompanied by substantial decrease in the effective local magnetic moment[27]. In several of these aforementioned reports, a direct comparison with the experimental curve is missing and show rather a reduced magnetization behavior. For instance, a study by *Lichenstein et.al.* shows the reduced ordered moment behavior with reduced temperature ($T/T_c$), wherein, only two data sets are used to compare with reduced experimental magnetization behavior[16]. The interpretation of results becomes ambiguous with low number of data points. Moreover, the accurate account of quatitative magnetization variation with temperature is equally crucial in order to exploit the material for technological applications and simultaneous validation of the theory for its predictability. In a recent study, *Hausoel et.al.*[28] provide a detailed analysis of self-energy and temperature dependent quasiparticle scattering rate, suggesting the presence of non-Fermi liquid behavior[28]. This study becomes important as they also show the presence of kink-like feature in the temperature dependent magnetic susceptibility curve. Interestingly these features have never been mentioned in the prior studies on Ni. This study differs from the earlier ones for having used first-principles calculated $U$ ($\sim$ 4 $eV$) and $J$ (1.08 $eV$) values obtained from constrained random phase approximation (cRPA). However, their characterization of the raised intensity region $\sim$ 6-8 $eV$ as the 6 $eV$ satellite feature is difficult to be quantified for the experimental 6 $eV$ satellite. Nonetheless, their results show $T_c$ (600 K) in quite good agreement with observed value (631 K) but the obtained saturation magnetic moment seems remarkably underestimated even with Full type of Coulomb interaction. It is essential to note that in majority of the studies, the electronic properties are obtained in the paramagnetic phase of Ni, where capturing the spectral features below $T_c$ would be challenging and might lead to misleading conclusions. Thus, the realistic approach would be considering the spin-polarization because the electronic structure of ferromagnetic $3d$-transition metals in the vicinity of the Fermi energy is dominated by spin-polarized $3d$ bands. Further, the 6 $eV$ satellite feature present below the chemical potential in the photoemssion spectra of Ni was largely identified to be correlation induced and significantly spin-polarized. We note that in the available studies a direct comparison of the calculated EES with the experimental EES is clearly missing which is essential to regard the calculated feature as the 6 $eV$ satellite by accounting for its observed intensity. Therefore, it becomes necessary to revisit its EES for better understanding of the underlying nature of its spectral features, especially satellite feature which can have multiple origins[29,30].

Further, it is noteworthy that certain experimental features of Ni have not yet been accounted in the available works, such as, the occurrence of satellite feature on slightly higher binding energy ($\sim$ 10 $eV$) in the photoemission spectra besides the 6 $eV$ satellite, which is also visible in the resonant photoemission spectra[31]. This satellite can be of particular interest due to its near proximity to 6 $eV$ satellite which represents a two-hole bound state[32]. Additionally, there have been mentions of temperature dependent contributions to the uniform magnetic susceptibility in experimental magnetization studies[33]. As stated before, DFT+DMFT method has been widely used to study Ni with a variety of implementations including different self-consistency schemes (eg. charge-self consistent DMFT) and different choices of impurity solvers[34–38]. It therefore becomes important to use the realistic formulation for better quantification of the observed properties.

Furthermore, the choice of $U$ and $J$ parameters is crucial to solve the correlated model Hamiltonian for the system's properties. Note that $U$ values used in most of the earlier studies lie in the range of 2-3 $eV$ ($J$$\sim$0.8 $eV$) typically taken from earliest of the reports where Coulomb interaction parameters were largely determined semi-empirically based on the good fit of either the electronic or magnetic properties. It would generally be poor practice to take $U$ and $J$ values from literature when the work does not use the same implementation of DFT+DMFT. This is especially the case for $U$ value chosen for the property specific case, which would typically not be adequate for the study of other physical properties. The first principles techniques developed to determine the Hubbard $U$ self-consistently include linear response theory[41], constrained Density functional theory (cDFT)[42] and constrained random phase approximation (cRPA)[43].

In these procedures the $U$ value is sensitive to the material specific parameters, including its position in

the lattice and the structural and magnetic properties of the crystal. In addition, it also depends on the choice of localized basis set and energy window employed to describe the on-site occupation in the correlated orbitals. The Coulomb interaction parameters can thus be self-consistently determined in accordance with the implementation of DFT+DMFT method, making the determination of $U$ and $J$ fully *ab initio*. Few of the later studies on Ni, which used the *ab intio* calculated $U$ within cRPA or cDFT were not found to exhibit large deviation from the earlier calculated EES, though the values of $U$ (3-6 $eV$)[27,28,44] were substantially higher than the ones used in literature (2-3 $eV$)[16,24,45]. However, the characterization of Ni based on the value of $U$ to classify it in a particular regime of correlation strength (strongly correlated vs moderately correlated) would be difficult as wide range of $U$ values have been used across the studies. Hence, on a single footing of Coulomb interaction parameters, $U$ and $J$, the account of the electronic and magnetic properties is crucial for the material's primary classification based on the strength of electronic correlation effects (identified through $U/W$ ratio; where $W$ stands for $3d$ bandwidth). We note that, quantitative description of both the magnetic and the spectral properties, with the same set of $U$ and $J$ parameters is lacking and has not yet been addressed completely. The real test of any theoretical framework lies in quantitative agreement of its results with experimental findings. The lack of agreement between the available results in quantitative respect using the advanced theoretical framework necessitates the need for a comprehensive study to account for a complete picture within the available computational resources.

In this paper, we have investigated both the magnetic and electronic properties of Ni within charge self-consistent DFT+DMFT approach through a comparative DFT and DFT+DMFT study, along with validating the suitable choice of Coulomb interaction parameters and form of Coulomb interaction paramterization to establish the general approach for correlated magnetic materials. We use cRPA instead of cDFT due to reports of unnecessarily high values of Coulomb interaction parameters obtained in cDFT[46]. This work includes first-principle calculations of $U(\omega)$ and $J(\omega)$ using cRPA[47]. We provide a self-consistent temperature dependent theoretical approach for the finite temperature magnetic properties' calculations witin DFT framework for better account of temperature effects for a comparative finite-temperature static mean-field and dynamical mean-field study. Interestingly, our results are indicative of better description of $M(T)$ behavior of Ni through the self-consistent account of temperature effects in comparison to DFT+DMFT approach. We also find a peculiar flat band (dispersion less feature) arising in the energy-dispersion curve largely dictating the finite temperature magnetic properties of Ni. Subsequently, exchange enhanced Pauli susceptibility contribution is noted in the paramagnetic spin susceptibility estimates. A leading Curie-Weiss behavior is eventually extracted at elevated temperatures. Our work presents a detailed analysis of the calculated EES followed by the reinterpretation of 6 $eV$ satellite alongside its temperature dependent spin-polarization estimates. Notably, significant plasmonic excitation contribution is found to the intensity of 6 $eV$ satellite from $G_0W_0$ results. Hence, providing a refined understanding of the famous satellite feature which was solely attributed to the electronic correlation effects in the past.

## II. COMPUTATIONAL DETAILS

In this work, the electronic structure calculations are carried out for Ni, wherein augmented plane wave plus local orbitals (APW+lo) method is used to carry out the DFT calculations with PBE exchange functional[48] using WIEN2k code[49]. The volume-optimized lattice parameter value of 3.513 Å is used with space group of 225. The convergence criteria for total energy is fixed at $10^{-4}$ $Ry/cell$ for 28 × 28 × 28 k-mesh. DFT+DMFT calculations are carried out with eDMFTF code[47] wherein, the single-particle Green's function is expanded in LAPW basis, and is fully charge self-consistently determined. WEIN2k performs the DFT calculation throughout DMFT iterations. Continuous-time QMC impurity solver is used with 'exactd' double-counting scheme[50]. Note that eDMFT implementation differs from other DFT+DMFT flavors. It describes all the valence states in the LAPW basis, which are allowed to hybridize with the correlated localized subset, unlike other implementations where itinerant states are approximated within tight-binding approximation. Subsequently, maximum-entropy method is used which brings the imaginary time/frequency calculated values of observables to real-time/frequency. The value of $U$ is calculated using cRPA by GAP2 code[51], where maximally localized Wannier basis function is used. The $J$ parameter is computed via Yukawa screening method[50] which proves to be a suitable choice for strongly correlated transition metals[52].The calculated value of $U$=5.78 $eV$ and $J$ (Hund's $J$)=1.1 $eV$ are used for the DFT+DMFT calculations where rotationally invariant type of Coulomb interaction paramterization (Full type) is considered. Further, Elk code[53] is used to perform temperature dependet $G_0W_0$ calculations for plasmon-frequency and electronic TDOS calculations.

.

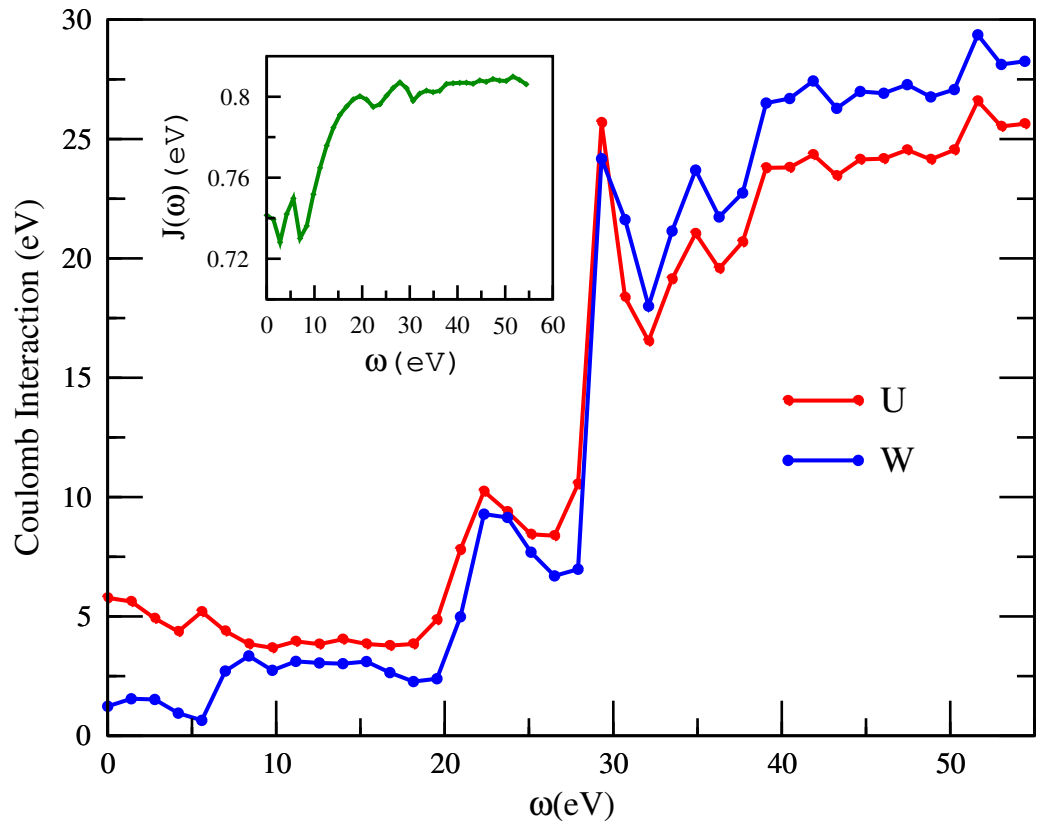


FIG. 1: Coulomb interaction as function of $\omega$ for Ni $3d$ orbitals

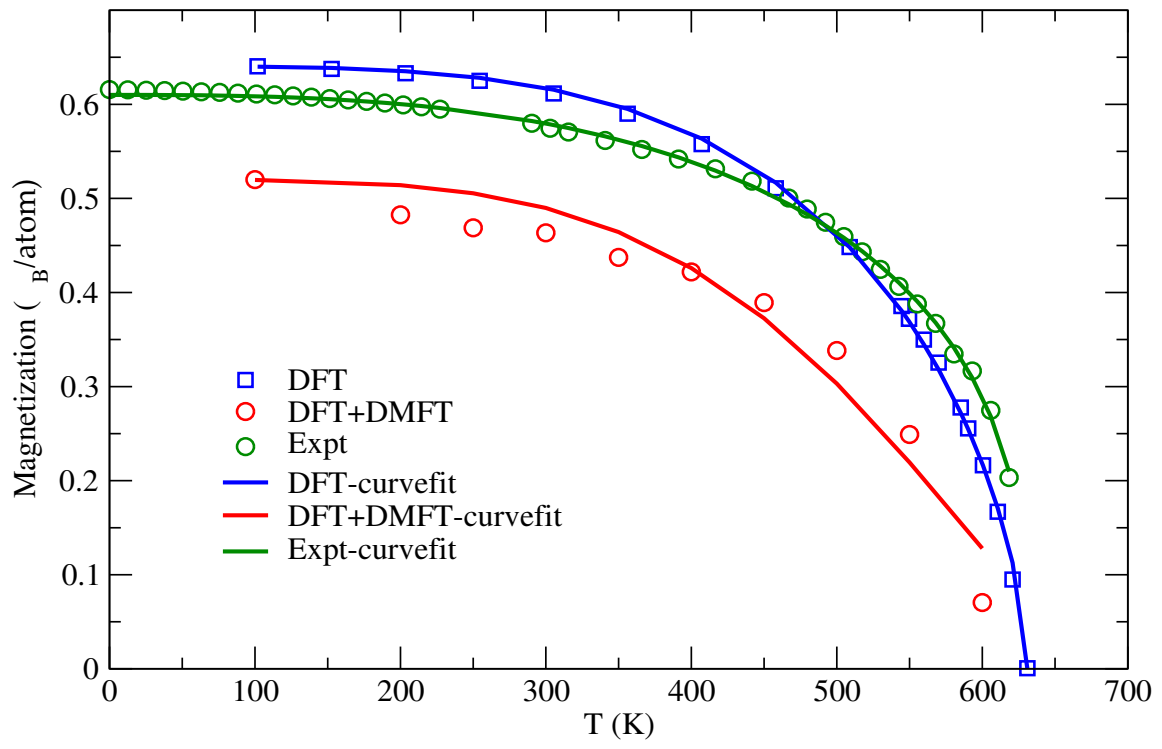


FIG. 2: Temperature dependent magnetization behavior calculated using DFT and DFT+DMFT methods along with experimental curve[56] with their best fit curves.

## III. RESULTS AND DISCUSSION

### A. Determination of Coulomb interaction parameters $U$ and $J$ using cRPA

The energy ($\omega$) dependence of partially screened Coulomb interaction parameter ($U$) and fully screened on-site Coulomb interaction parameter ($W$) along with Hund's exchange parameter ($J$) are given in Fig. 1. The values of $U$, $W$ and $J$ are calculated from cRPA by taking the average value of all matrix elements of their respective Coulomb interaction matrices, excluding the screening window of bands having predominantly $3d$ character. The extent of screening can be inferred from the value of partially screened $U(\omega{=}0)$ (5.78 $eV$) which is found to be $\sim$23% of its bare value ($\sim$24.9 $eV$). This suggests the importance of screening effects in computing the material specific $U$. However, $J(\omega)$ remains nearly constant and close to its bare value ($\sim$0.84 $eV$). Which is consistent with the fact that $J$ is weakly sensitive to the screening effects in solid and is found to be close to its atomic value ($\sim$0.8 $eV$).

The sudden drop in $W$ value around $\omega = 5.7\ eV$, marks the onset of plasmon excitations[52,54]. Since, DMFT formulation already includes screening among the correlated orbitals, therefore partially screened on-site Coulomb interaction parameter $U$ (obtained by excluding the $3d$-$3d$ band polarization) is used instead of fully screened (generally referred to as $W$), as the $U$ parameter in DFT+DMFT calculations. Here, the value of $U$ obtained turns out to be higher than some of the earlier cRPA estimates of $U$= 2.8 $eV$[46], 4.0 $eV$[55], where not much clarity is present on the kinds of constraints, tweaking parameters and the bands selected, in order to compare any such results on relative grounds. In our case, although the band weight scale option (a tweaking parameter) can be used to tailor $U$ to further low value, but we find it essential to consider the results without any additional parameter dependence for the genuine representation of a material's attributes.

### B. Finite-temperature magnetic properties

**$M(T)$ curve:** Fig. 2 shows $M(T)$ curves calculated using DFT and DFT+DMFT approaches along with the experimental curve. We realize the need to revisit finite temperature magnetization behavior not only with the aim of studying $T_c$ but to simultaneously understand the origin of the magnentism (being dictated predominantly by spin-fluctuations or itinerancy in Ni $3d$ states) through a comparative DFT and DFT+DMFT study. Since, finite-temperature excited states cannot be studied in DFT explicitly, we have obtained the $M(T)$ curve for Ni, within the DFT framework by including the temperature dependence implicitly via temperature dependent occupation number/ occupancy of states. This is achieved by employing the Fermi-dirac distribution function in each iteration until the convergence is reached. We note that DFT gives a saturation magnetization of 0.64 $\mu_B$/Ni, in close agreement with the experimental saturation magnetization (0.61 $\mu_B$). However, as expected the collapse of the total magnetic moment takes place at very high electronic temperature ($T_{ec}$) of 3100 K, consistent with the previous mentions of strikingly high critical temperature obtained using mean-field studies on Ni[2]. This deviation was largely ascribed to the influence of spin fluctuations dominating the description of $T_c$ in these materials[3]. The DFT+DMFT curve consistently underestimates the magnetization values over the given temperature range with relatively low saturation magnetization value (0.52 $\mu_B$/Ni) and slightly underestimated critical temperature ($\sim$ 600 K).

Nevertheless, we observe that scaling the electronic temperature range in DFT calculations with a scaling factor obtained as the ratio of $T_c/T_{ec}$ ($\sim$0.2) (where, $T_{ec}$ = 3100 K is the critical electronic temperature obtained from DFT method, $T_c$ = 600 K is the magnetic transition temperature obtained from the DFT+DMFT calcu-

lations) results in an exceptionally good match with the experimental $M(T)$ behavior. This suggests the overestimation of the electronic temperature by a factor of ∼5 for Ni and correspondingly the thermal energy needed to collapse the magnetic moment within the DFT approach. Which can be attributed to the absence of spin-fluctuations that would essentially lower the critical temperature further. Also, this approach can be used as a potent tool to extract the observed $M(T)$ behavior, wherein material specific temperature scaling factor ($T_c/T_{ec}$) can be easily determined from the DFT+DMFT obtained $T_c$. This result for Ni underscores the predominance of itinerant nature of $3d$ electrons over the spin-fluctuations in deciding its $M(T)$. The $M(T)$ curve obtained from static mean-field approach has also been a subject of interest in studying the ultrafast demagnetization dynamics in $3d$ transition metals[57].

It would be interesting to study the applicability of this approach in case of other transition metals.

Besides, the notable suppression of magnetization values in DFT+DMFT method can be understood in terms of the overestimation of spin-fluctuations throughout the temperature range especially in low temperature region. This might be due to the additional account of quantum spin fluctuations along with thermal spin fluctuations. Another plausible reason could be assigned to the neglection of non-local correlations in single site DMFT approach, which only describe the dynamical local correlations via k-independent self-energy.

The two fundamental models in literature for describing the nature of finite-temperature magnetism in materials were typically derived based on the (i) single particle excitations (Stoner model) and (ii) spin-fluctuations[2]. In Fig. 2, we have shown the best-fit curves obtained from both the DFT and DFT+DMFT methods along with the best fit for experimental curve to compare with the fitting equation (Eq. 1) of aforementioned magnetism models *i.e.* Stoner model and spin-fluctuation model.

$$M(T) = M_0(1 - (T/T_c)^m)^n \tag{1}$$

Where $M_0$ and $T_c$ are the saturation magnetization and $T_c$ values respectively, for the corresponding methods. The standard values of parameters $m$ and $n$ are 2 (1) and 0.5 (0.5) for the stoner model (spin-fluctuation model). The best fit values of parameters obtained for the calculated and experimental curves are given in table I.

TABLE I: The value of parameters $m$ and $n$, obtained corresponding to the best fit experimental and DFT & DFT+DMFT, $M(T)$ curves.

| **Parameters** | **DFT** | **DFT+DMFT** | **Expt** |
|---|---|---|---|
| $m$ | 3.83 | 4.06 | 2.73 |
| $n$ | 0.61 | 1.83 | 0.36 |

In this case, both the models are found to fail to describe the calculated and experimental curves. The failure of both the models to depict the experimental behavior shows that the nature of magnetism in Ni cannot be classified to be predominantly coming from either the itinerant nature of $3d$ electrons or spin-fluctuations.

**spin susceptibility:** The dynamical local spin susceptibility computed from the CT-QMC calculations at zero frequency, as given by Eq. 2[16], is shown in Fig. 3 in the temperature range of 700-3100 K.

$$\chi_{loc}(\omega = 0) = \frac{g_s^2}{3}\int_0^\beta \langle \mathbf{S}(\tau)\mathbf{S}(0)\rangle d\tau \tag{2}$$

Where $\beta$ is the inverse temperature, $g_s$=2 is the electron spin gyrometric ratio, and $\langle S_z(\tau)S_z(0)\rangle$ is the imaginary time-dependent spin-spin correlation function. The $\chi_{loc}(\omega = 0)$ denotes the static spin susceptibility in the local-moment regime. Its temperature dependence holds utmost importance in characterizing the degree of correlations in the system. In weakly correlated systems $\chi_{loc}$, is expected to be almost temperature independent; whereas, in the presence of strong correlations, a dominating Curie-Weiss behavior is expected at high temperatures. The calculated $\chi_{loc}$ is given in Fig. 3(a). The gradual decrement in $\chi_{loc}$ values above T∼2000 K, indicates clear deviation from the Curie law. Typically, the $\chi_{loc}$ computed using DFT+DMFT includes various competing many-body effects, such as temperature induced local moment formation, disordering of local moments by thermal fluctuations, screening of local moments by conduction electrons, etc[3,58]. Here, we find the deviation of $\chi_{loc}$ from the $1/T$ law due to the presence of Pauli-like susceptibility contribution dominating at higher temperatures. This is apparent from the enhanced curvature of the $\chi_{loc}$ after subtracting the Pauli-susceptibility ($\chi_p$ ) contribution (Fig. 3(a)). Note that $\chi_p$ in its theoretical definition is majorly a temperature independent contribution, which is given in Eq. 3.

$$\chi_{pauli} = \mu_B^2 g(\epsilon_f) \tag{3}$$

where, $g(\epsilon_f)$ is the free electron density of states at the Fermi-energy ($\epsilon_F$). According to the theory, it represents response of non-interacting electrons in the presence of external magnetic field. It is thus defined purely depending on the non-interacting total density of states at the $\epsilon_F$. The nearest theoretical framework for the description of the non-interacting particle picture is the DFT approach excluding the exchange correlation potential. Thus the $\chi_p$ contribution calculated from the total density of states is found to have notably high value of ∼ 1.62 $\mu_B^2/eV$.

The dominating Pauli contribution to the total spin susceptibility is also discussed in earlier experimental reports. The mentions of Pauli susceptibility contribution estimated from low-temperature electronic specific heat coefficient in ferromagnetic phase are found to be

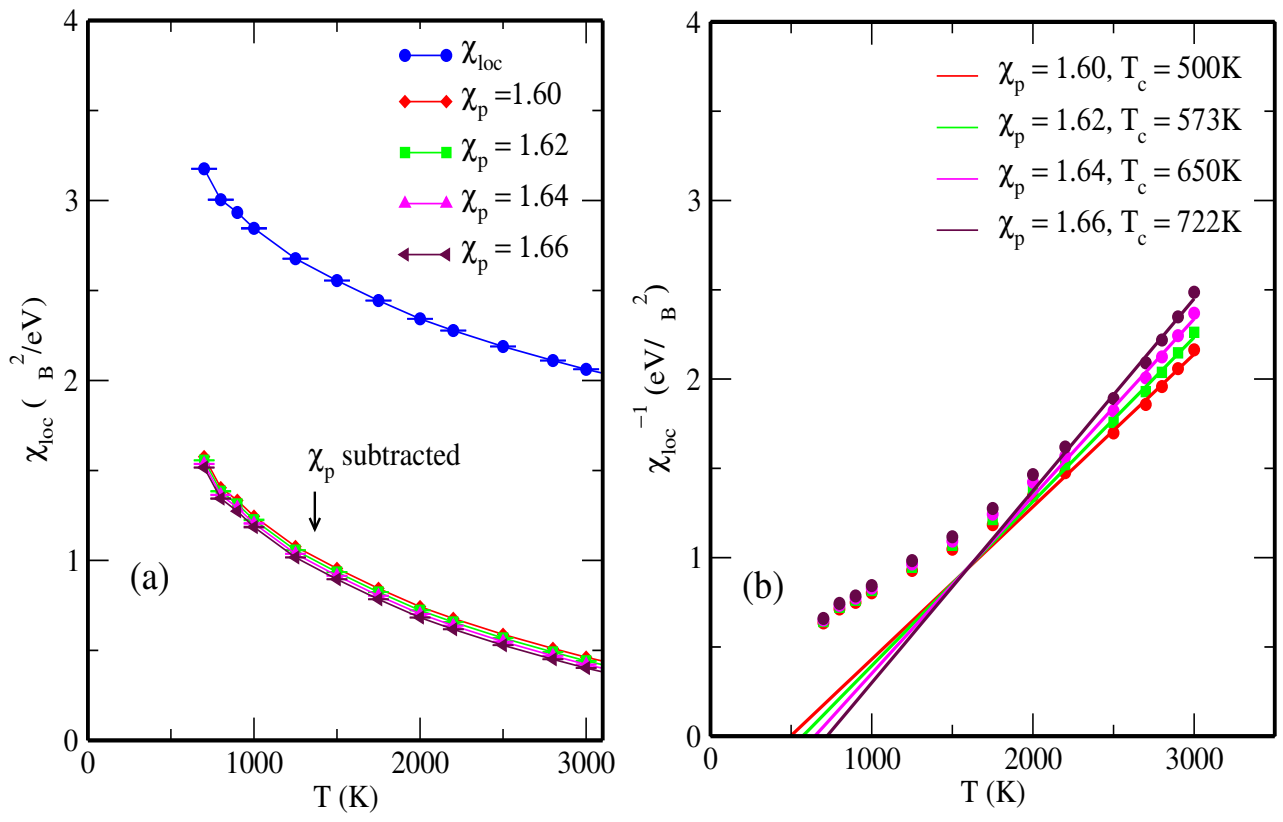


FIG. 3: DFT+DMFT calculated local spin-susceptibility (a) Shows total local magnetic susceptibility ($\chi_{loc}$) along with $\chi_p$ subtracted local magnetic susceptibility curves corresponding to four different values of $\chi_p$ as shown. (b) Shows $\chi_p$ subtracted inverse local magnetic susceptibility curves along with their respective Curie-Weiss fits and obtained $T_{c_s}$ .

significant[59,60]. The Fig. 3(b) shows inverse $\chi_{loc}$ with the best linear Curie-Weiss fit corresponding to slight variation in $\chi_p$ values for extracting the observed $T_c$. We find remarkable dependence of $T_c$ on the value of $\chi_p$. This is directly evident from considerable change in slope of $\chi_{loc}$ with the mentioned variations in subtracted $\chi_p$ contributions and ultimately reflects in large variation in correspondingly calculated $T_c$ values from the Curie-Weiss fit (Fig. 3(b)). The difference in $T_c$ values by 150 K with a mere 0.04 $\mu_B^2/eV$ change in $\chi_p$ suggests the importance of accurate determination of $\chi_p$ for extracting the observed $T_c$ of Ni.

Further, temperature dependence of uniform magnetic spin susceptibility ($\chi$) and $\chi_p$ is evaluated using finite-temperature DFT method. The Eqs. 4, 5 and 6 depict the self-consistent temperature dependent DFT framework employed to extract finite-temperature $\chi$ and $\chi_p$ values.

$$M(H,T) = \frac{1}{V}\sum_i n^f_{i_\uparrow} - n^f_{i_\downarrow} \tag{4}$$

$$\chi(T) = \frac{\delta M(H,T)}{\delta H} \tag{5}$$

where $i$ represents the $i^{th}$ state and $n_i^f$ denotes the finite temperature occupation number of $i^{th}$ state with $n^f_{i_\uparrow}$ ($n^f_{i_\downarrow}$) referring to the occupancy of spin-up (spin-down) states at temperature $T$.

The finite temperature occupation for respective states are calculated using the Fermi-Dirac distribution function.

$$n_i^f(T) = \frac{1}{e^{(\frac{\epsilon_i(T)-\mu}{k_B T})} + 1} \tag{6}$$

where $\epsilon_i(T)$ represents the energy eigenvalue of the $i^{th}$ state at temperature $T$ with correspondingly determined chemical potential $\mu$ and $k_B$ is the Boltzmann constant. These equations thus depict the employment of temperature dependent framework (consisting of calculation of excitation energies of states at finite temperature $\epsilon_i(T)$ and thus $\mu$ in self-consistent manner). This approach is expected to be more accurate finite temperature material description within DFT. Note that the estimation of conventional $\chi_p$ is made possible via solving the Kohn-sham Hamiltonian in DFT, excluding the exchange correlation potential and calculating the change in magnetization in the presence of few magnetic field variation to ensure the linearity of response. The $\chi_p(T)$ curve thus obtained is shown in Fig. 4. Since, the $T_c$ value in DFT calculations is found to be highly overestimated by the order of ∼5 ( $T_c$ ∼3100 K; experimental $T_c$ ∼631 K), the calculations of $\chi$ and $\chi_p$ are thus carried out in the corresponding paramagnetic phase above 3100 K. We note significant temperature dependence of $\chi_p$ over the given temperature range, for instance ∼36% change in its value over 4500 K difference of temperature.

Fig. 5(a) shows the temperature dependence of $\chi$ ($\chi$(T)) along with temperature dependent $\chi_p$ ($\chi_p$(T)) subtracted curve. A clear deviation of $\chi$(T) from the Curie law ($1/T$) could be seen, which improves upon subtracting the $\chi_p(T)$ contribution. Also, the inset of Fig. 5(a) closely shows the enhanced curvature of $\chi$(T) upon $\chi_p$(T) subtraction, leading to improved agreement with $1/T$ behavior. The diverging value of $\chi$(T) ∼ 3100 K is consistent with the fact that $\chi$ diverges at $T_c$ (since collapse of magnetic moment is observed ∼ 3100 K within DFT). Subsequently, the $\chi^{-1}$(T) shows improved slope of linear fit in agreement with the Curie-Weiss law, given in Fig. 5(b). The Table. II shows the Curie-Weiss fit of $\chi^{-1}$(T) and obtained $T_c$ values in certain high temperature regions with $T_{c1}$ and $T_{c2}$ being the critical temperature values obtained before and after subtracting the $\chi_p$(T) contribution, respectively. Moreover, the table shows improved description of critical temperature values after ruling out the $\chi_p(T)$ contribution irrespective of the fitted temperature range considered (For reference see $T_{c2}$ in Table II.). However, the best description of $T_c$ is found for the Curie-Weiss fit considered over the whole temperature range (*i.e.* 3100-8000 K; $T_{c2}$∼3188 K). Here, the $T_c$ (∼3188 K) obtained from the Curie-Weiss fit of $\chi^{-1}$(T) across a broad temperature range comes out to be close to the $T_c$ observed from the collapse of magnetic moment from the DFT $M(T)$.

The above DFT and DFT+DMFT results clearly show significant contribution of both $\chi_p$ and Curie-Weiss susceptibility to the total paramagnetic spin susceptibility. This shows the presence of competing degrees of both itinerancy (inferred from high $\chi_p$ contribution) and localized moments (evident from the Curie-Weiss fit), dictating the observed finite-temperature magnetic properties of Ni. Now we discuss the plausible reason for the large Pauli-spin response of the magnetic susceptibility in Ni. It is found to be arising from a peculiar flat band feature present in close proximity of the $\epsilon_f$ along $X$-$W$ direction in the energy dispersion curve. It can be more illustratively explained based on the DMFT calculated k-resolved spectral function plot. Fig. 6 depicts the k-resolved spectral function plot at 300 K which shows the presence of flat band feature migrated at the $\epsilon_F$ when compared to FM DFT bandstructure (given in Fig. 7). Clearly, the temperature effects and inclusion of dynamical corrections to the single-particle energy eigenvalues leading to large renormalizatin of single-particle energy band structure is evident. It is found to be situated almost at the $\epsilon_F$ in k-resolved spectral function plot across broad temperature range and thus, being responsible for the enhanced $\chi_p$. We suspect the temperature-dependent flat band feature extending along the $X$-$W$ direction, largely influencing the finite-temperature magnetic properties of this system. More recently, the effect of such $3d$ flat bands in the vicinity of $\epsilon_F$ of CoSn has been studied with hole or electron doping for tuning its flat band position with respect to $\epsilon_F$ for enhanced Pauli-susceptibility and potential applications[61]. Our result also provides explanation to one of the earliest works on Ni rich Ni-Nb alloy, where enhanced temperature dependent contribution to paramagnetic susceptibility is noted with increase in Nb concentration. The Nb concentration increase can be seen equivalent to hole-doping, thus driving the flat band towards $\epsilon_F$ and gives rise to the increased temperature dependent part of the magnetic susceptibility.[62].

Unlike our result, *Hausoel et.al.*[28] show a kink like feature in their local spin susceptibility curve in both non-interacting and interacting Coulomb interaction parameterizations. They attribute the presence of kink at temperature $\sim$ 1200 K corresponding to the position of flat band feature from the $\epsilon_F$, calcualted at $T = 0$. However, as mentioned above, we find the flat band position with respect to $\epsilon_F$, to be largely temperature dependent. Which suggests a gradual shift in the position of this flat band feature towards the $\epsilon_F$. This gradual shift would lead to finite increase in occupancy of the flat band which implies gradual increase in the finite-temperature occupancy of the flat band states. Hence, it is not expected to manifest as sudden enhancement of local moments or the kink, as suggested by our results. Moreover, this kink like feature is not ubiquitous in the experimental spin susceptibility data, as it is so far reported in one experimental study[63].

In order to further validate the presence of local moments, we have also calculated fluctuating moments ($\langle S_z^2 \rangle$) at finite temperature. The $\langle S_z^2 \rangle \sim 0.40$, is found to remain almost constant in the studied temperature range (200-3000 K). The temperature independent behavior of $\langle S_z^2 \rangle$ suggests the presence of local moments in

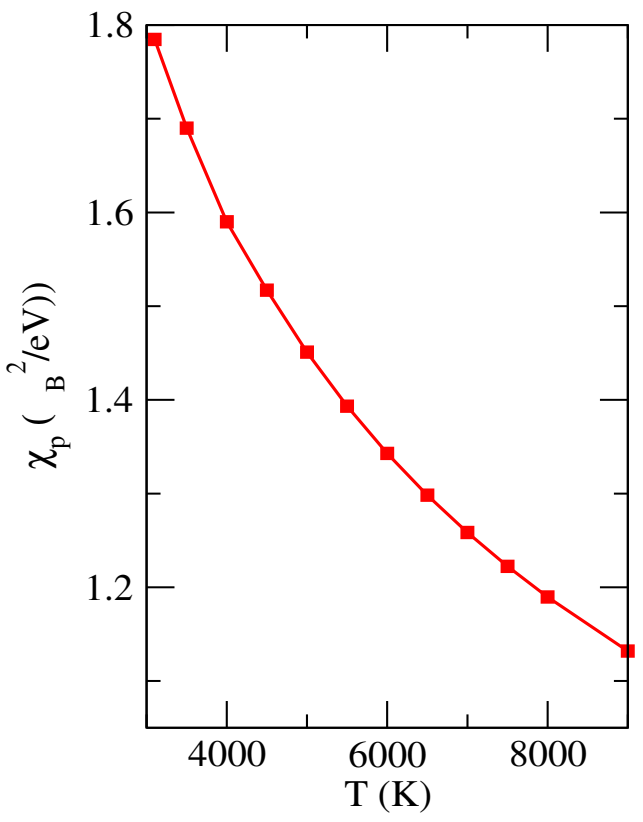

FIG. 4: DFT calculated temperature dependence of Pauli spin suscpetibility ($\chi_p$)

Ni. Also, the probability of electronic configuration of Ni $3d$ orbitals is calculated and comes out to be temperature independent. The electronic configuration state probabilities calculated at 300 K are depicted in Fig. 8. The results show a mixed-valence state electronic configuration probable for Ni, with $3d^8$ ($\sim$0.41) and $3d^9$ ($\sim$0.40) seem to be relatively equally maximally probable and is consistent with the experimental findings of corresponding degenerate ground state electronic configurations[64]. Interestingly, $3d^7$ ($\sim$0.11) is also found to carry significant probability of occupation. From the figure (Fig. 8), charge fluctuation $\delta N = \langle (N - \langle N \rangle)^2 \rangle$ is further evaluated and is found to remain constant across the temperature range ($\delta N \sim 0.67$). The appreciable magnitude of charge fluctuations suggest a significant degree of itinerancy present in the system. Thus, the above results exemplify the comparative role of both the degree of itinerancy and localization of $3d$ electrons, in deciding both the electronic and magnetic properties of Ni.

TABLE II: Critical temperature values obtained from the Curie-Weiss fit of the total uniform static susceptibility ($T_{c1}$) and Pauli paramagnetism contribution subtracted uniform static susceptibility ($T_{c2}$) curves in different considered temperature range to see the dependence of the estimated $T_c s$ on the consideration of temperature window for the Curie-Weiss fit. Temperature range (T), $T_{c1}$ and $T_{c2}$ are given in the units of Kelvin (K).

| **T range** | $T_{c1}$ | $T_{c2}$ |
|---|---|---|
| 3100-8000 | 2739 | 3188 |
| 4000-8000 | 2442 | 3305 |
| 5000-8000 | 2124 | 3400 |

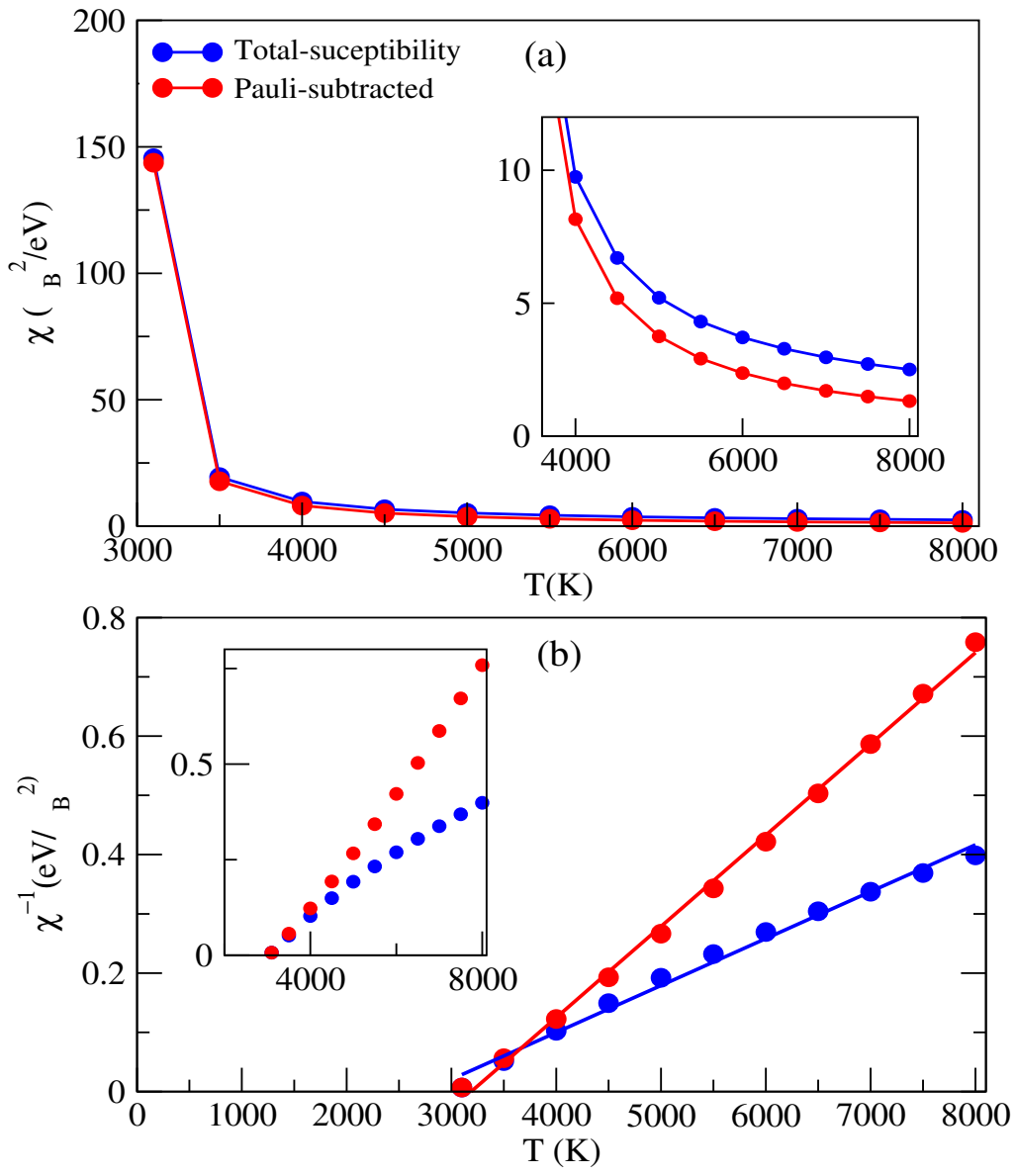

FIG. 5: DFT calculated paramagnetic (a) uniform magnetic spin susceptibility ($\chi$) showing both total susceptibility and Pauli susceptibility ($\chi_p$) subtracted curves (b) Inverse uniform magnetic total susceptibility and Pauli subtracted inverse uniform magnetic susceptibility curve with their corresponding best linear fits

### C. Spectral properties study

The DFT+DMFT study shows significant renormalization of the Ni single-particle energy-bands especially in the proximity of $\epsilon_F$, which can be seen from the $k$-resolved spectral function plot (at $T$=300 K) calculated using DFT+DMFT (Fig. 6) and band-structure plot obtained from the DFT approach (Fig. 7). Remarkable change in the energy position of the bands after employing DMFT correction, indicates sign of strong correlation effects in the system. The $k$-resolved spectral function plot shows the presence of coherent and incoherent band features (where the color scale adjacent to the figure depicts increasing degree of coherency in the energy-band moving from red(min) to yellow(max)).

The temperature dependence of inverse quasiparticle lifetime $\Gamma$ ( quasiparticle scattering rate) in non-magnetic (NM) and ferromagnetic (FM) phases, is given in Fig. 9(a) and Fig. 9(b), respectively. The $\Gamma$ values for the respective $e_g$ and $t_{2g}$ states at each temperature, were calculated considering the imaginary part of self energy in the Matsubara frequency domain ($Im\Sigma(i\omega)$) to avoid any kind of noise participation due to the problems inherent to analytic continuation methods. Thus the expression of calculating $\Gamma$ sums up as follows: $\Gamma = -ZIm\Sigma(i0+)$); where $Z$ is the mass renormalization factor; $Z^{-1} = 1-\delta Im\Sigma(i\omega)/\delta\omega$. Interestingly, $\Gamma$ for both the $e_g$ and $t_{2g}$ states in NM phase, show an evident deviation from the $T^2$ behavior as predicted by the standard Fermi-liquid theory, thus showing sign of non-Fermi liquid behavior (NFL) (given in Fig. 9(a).). Which is consistent with the recent report on Ni[28]. However, we find no clear sign of the presence of NFL behavior in the $\Gamma$ calculated in the FM phase (Fig. 9(b)) in temperature range (200-600 K). We do not find theoretical studies showing the temperature dependence of $\Gamma$ in FM phase of Ni. The deviation from Fermi-liquid behavior could be attributed to the large renormalization of the flat band, leading to increased degree of incoherency and driving the $\Gamma$ variation to NFL behavior.

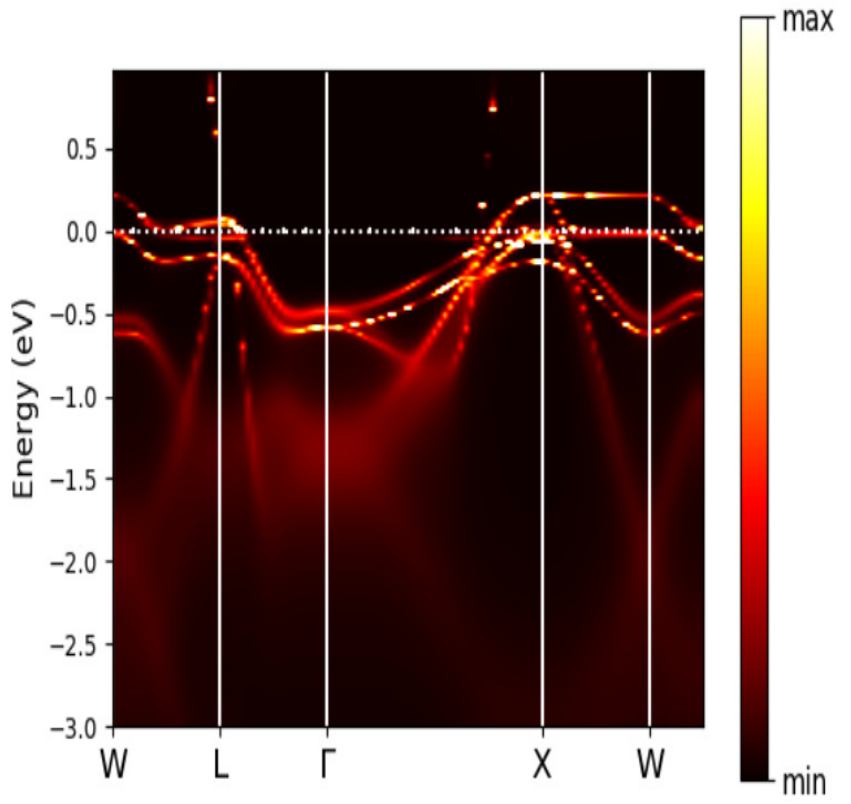

FIG. 6: The momentum-resolved spectral function plot of Ni at 300 K obtained using DFT+DMFT in FM phase; The color gradient adjacent to the graph shows the increasing degree of coherency from *min* to *max* as marked on the scale.

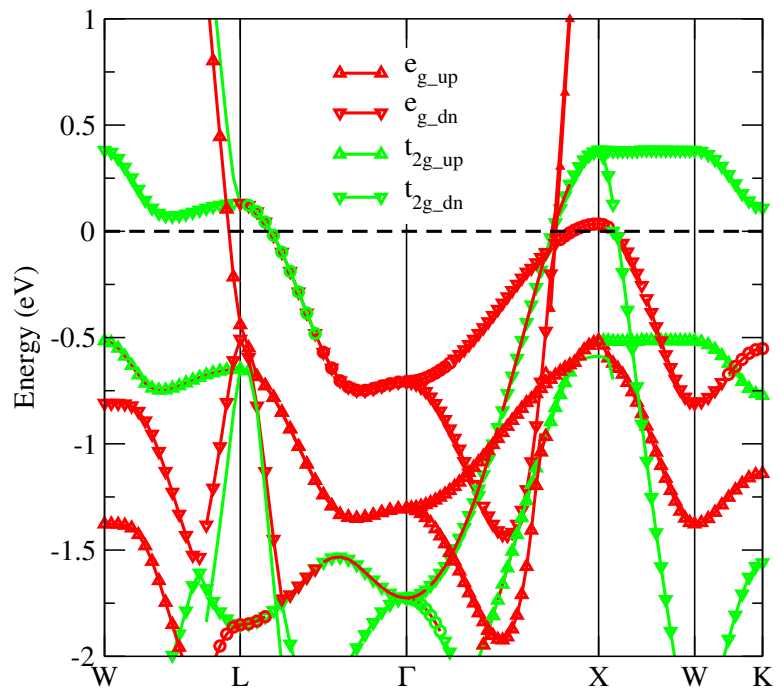

FIG. 7: The spin resolved single particle energy dispersion curve of $\epsilon_g$ and $t_{2g}$ states obtained within DFT framework. .

Fig. 10 shows the calculated EES using DFT and DFT+DMFT ($T$=300 K) methods along with the observed X-ray photoelectron spectroscopy (XPS) valence band spectrum of Ni[66]. The Mg K-$\alpha$ X-rays (1253.6 $eV$) source is used in obtaining the valence band spectrum of Ni. For this source, the cross-section of $3d$ states is found to be ~10 times that of $4s$ states[65]. Therefore, the primary contribution to the spectrum can be considered to come from the $3d$ states. So, the spectral function of $3d$ states is shown in the DFT+DMFT curve. For calculating the spectral function corresponding to occupied levels, the density of states (DOS) obtained from respective theoretical methods is convoluted with the instrumental broadening of 0.55 $eV$, on the top of Fermi-Dirac distribution correction to the states[32,66]. The failure of DFT in accounting for the experimental curve is found to be consistent with previous studies[16,28,67–69]. The deviation of DFT calculated EES, reflecting as broadened line width in comparison to experimental EES could be explained by the overestimated $3d$ bandwidth by almost 30% at DFT level[67]. The DFT+DMFT obtained EES provides an appreciably good match of line shape and line width with experimental EES, over the studied frequency range. The slight mismatch of intensity between the DFT+DMFT EES and experimental spectrum can be accounted considering the inherent corrections in the extraction procedure of photoemission spectra such as, background corrections, unknown background function and inelastic electron contributions, etc. However, the difference in intensity generated near the 6 $eV$ between the theoretical and experimental curve and cannot be attributed to the above mentioned limitations of the XPS measured spectrum. In agreement with the earlier reports, we also note ∼30% narrowing of the $3d$-band and substantial reduction in the exchange splitting energy, with the application of DFT+DMFT approach relative to DFT results[67–69]. We note the extension of a plateau like feature extending from 6 $eV$ till 8 $eV$ and above. Although, presence of this feature is not visible in the experimental curve shown, this can be ascribed to the fact that the experimental EES here is background subtracted, where region above 8 $eV$ is part of the background. The presence of additional satellite feature in the region ∼ 10 eV is consistent with the resonant photoemission data for Ni[31]. It is noteworthy that the value of $U$ (5.78 $eV$) used here is able to describe the other satellite feature ( ∼ 10 $eV$) as well. Which is otherwise not generated at smaller $U$ values as evident from the aforementioned literature studies. Besides, we have studied the temperature dependent spin-polarization of 6 $eV$ satellite. Its spin-polarization exhibits a drastic decrease from 14% to 4% in the temperature range ($T/T_c = 0.66$, 0.8, 1; $T_c$=600 K). This decrease in the spin-polarization value with increasing temperature, is consistent with the experimental results of *Kakizaki et.al.*[31].

The appropriateness of the Coulomb interaction parameters used here, stand validated, considering the improved description of the experimental EES within the limitations of XPS data interpretation, except for the 6 $eV$ satellite intensity. The presence of 6 $eV$ satellite feature in the photoemission spectra of Ni has been largely attributed to the electronic correlation effects of $3d$ states[70]. Nevertheless, the presence of 6 $eV$ plasmonic loss has also been proposed by energy loss spectrum studies on Ni[71,72]. Our study suggests that the 6 $eV$ satellite is not solely electronic correlation induced in nature but additionally contains the contribution from the plasmonic excitations for its experimental intensity account.

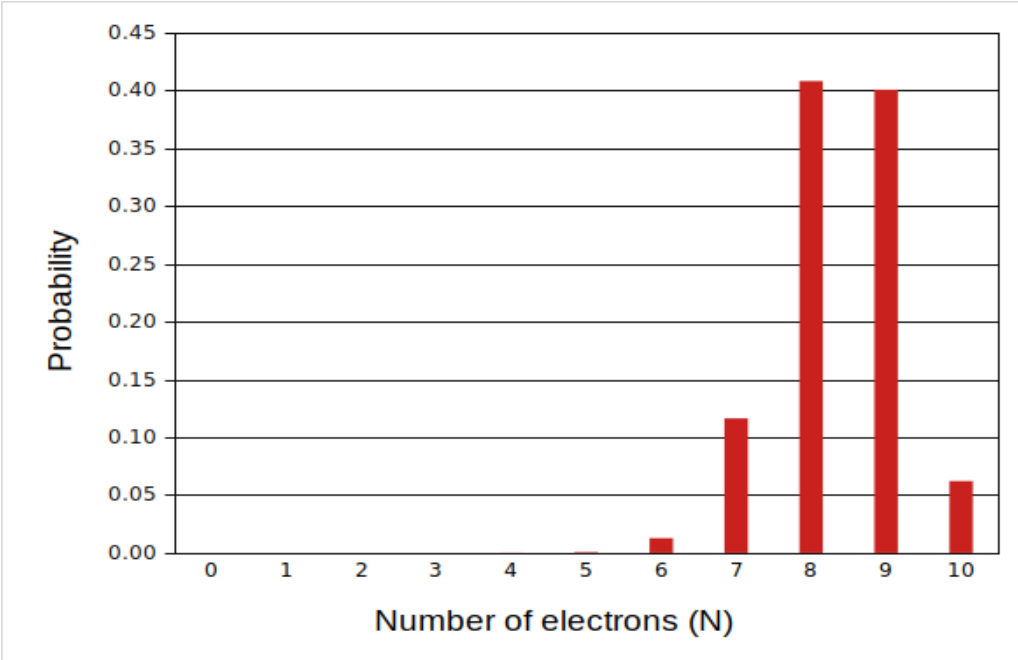

FIG. 8: Probability of atomic electronic configuration corresponding to N electrons in Ni $3d$ orbitals at 300 K. .

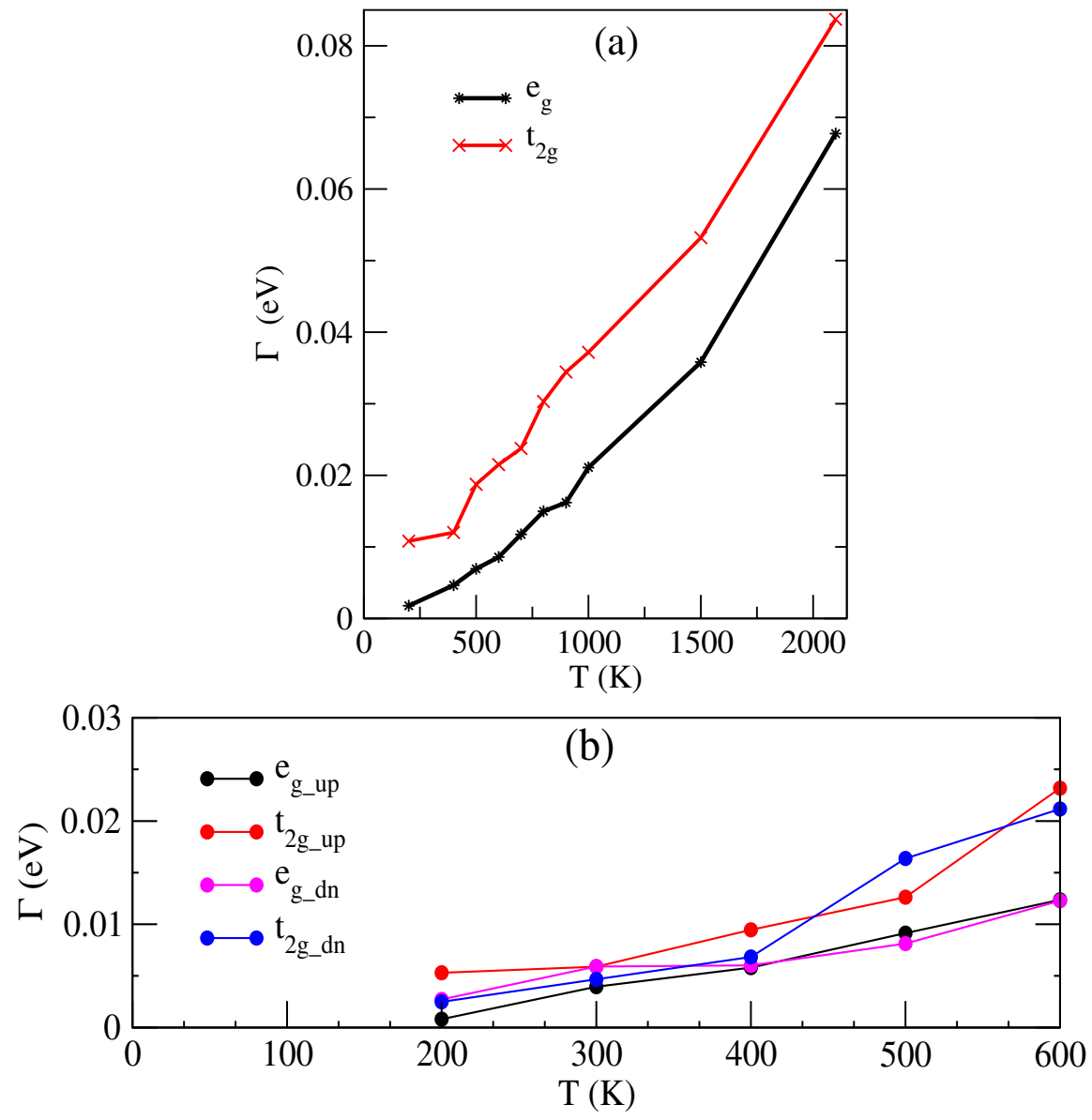

FIG. 9: The variation in inverse quasiparticle lifetime/ quasiparticle scattering rate (Γ) of $e_g$ and $t_{eg}$ states of Ni with temperature, in (a) NM phase; The variation shows deviation from the $T^2$ behavior as predicted by the Fermi-liquid theory. and (b) FM phase; The variation shows fluctuating behavior with temperature approaching $T_c$. .

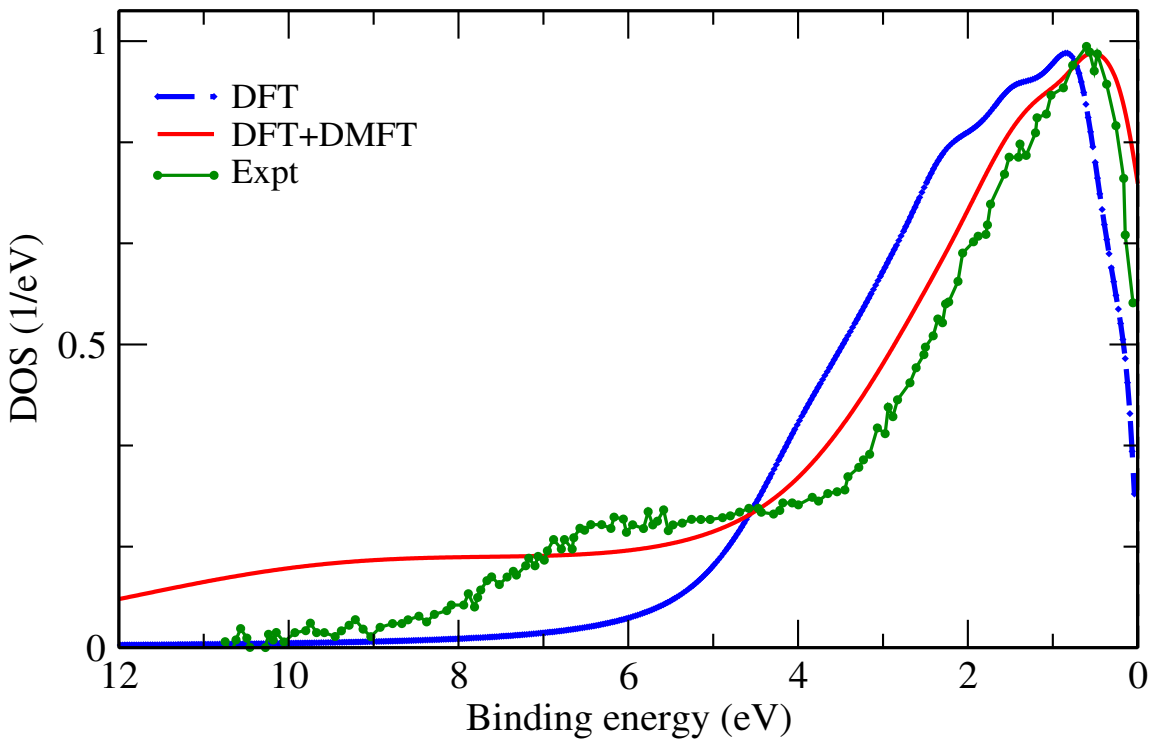

FIG. 10: Plot of theoretical EES for ferromagnetic Ni, obtained using DFT and DFT+DMFT ($T$= 300 K) methods, along with the experimental EES[66] curve.

### D. $G_0W_0$ results for plasmonic excitaions

Finally, we present the results of $G_0W_0$ calculations performed for the plasmonic excitation frequency and TDOS of Ni using random phase approximation (RPA). Fig. 11 shows the calculated DOS of the occupied band in both the NM and FM phases at 300 K. The parts (a) and (b) of the Fig. 11, clearly show the presence of satellite like feature ∼6 $eV$ in both its NM and FM phases, respectively. Since, the spectral weight due to plasmonic loss cannot be unambiguously identified from the k-integrated plot (DOS), we have thus shown the spectral function plot ($A_{k\omega}$) (at k-point coordinate (0.333, 0.166, 0.083)) corresponding to both NM and FM phases in Fig. 11((c) and (d)), respectively. This is done for the distinct depiction of spectral weight transfer mainly due to the renormalization of quasiparticle peaks by excluding the features or added spectral weights which are present in the k-integrated plot.

Also, note that the straight lines in Fig. 11(c) and 11(d) represent the DFT peaks, and the spectral function plot obtained after the renormalization of these DFT peaks into quasiparticle peaks in NM and FM phases, respectively. In NM phase (Fig. 11(c)), a weighted peak like feature having two shoulder peaks is obtained ∼ 6 $eV$ frequency which cannot be explained by the resultant quasiparticle peak corresponding to the DFT peak present slightly above 6 $eV$. It can be explained considering additional contribution to its spectral weight due to the spectral weight transfer from the quasiparticle peak around the $\epsilon_F$ (recognized as ∼6 $eV$ plasmonic loss). Similarly, the presence of largely broadened three-peaked structure ∼6 $eV$ in the FM phase (Fig. 11(d)) could be explained due to the collective contribution of spectral weight from the quasiparticle peaks corresponding to the exchange splitted DFT peaks closely spaced around 6.2 $eV$ and the spectral weight transfer from the quasiparticle peak situated around the $\epsilon_F$ (6 $eV$ plasmonic loss).

The presence of plasmonic excitations ∼6 $eV$ can also be confirmed through the calculation of plasmon frequency in both NM and FM phases. Also, note that in FM phase the 6 $eV$ feature extends till ∼6.5 $eV$ which is slightly large in comparison to its extension in NM phase. This can be explained from the difference in plasmon frequency ($\omega_p$) estimates obtained in both the phases. The plasmon excitation frequency in FM phase ($\omega_p$∼ 6.9 $eV$) is found to be higher than in the NM phase ($\omega_p$∼ 6.04 $eV$). Since the FM RPA calculations at finite-temperature are ill-defined, due to improper fluctuating moment account at finite-temperature, hence the real $\omega_p$ gets slightly overestimated. The $\omega_p$ in the NM case seems in close agreement with the frequency region of observed satellite-peak ∼6 $eV$, which serves as qualitative indicator of the presence of plasmon excitations' contribution to the 6 $eV$ satellite peak. Such a signature of the existence of plasmon-excitations ∼6 $eV$, is also evident from the $U(\omega)$ plot as previously mentioned in text (Fig. 1) wherein, remarkable dip in the value of on-site Coulomb interaction parameter is seen ∼ 6 $eV$ frequency region. Therefore, the appearance of 6 $eV$ satellite peak in the XPS obtained for Ni, can be regarded as a consequence of collectively two major contributions, *i.e.*, correlation-effects in $3d$ states and the plasmon excitations found ∼6 $eV$.

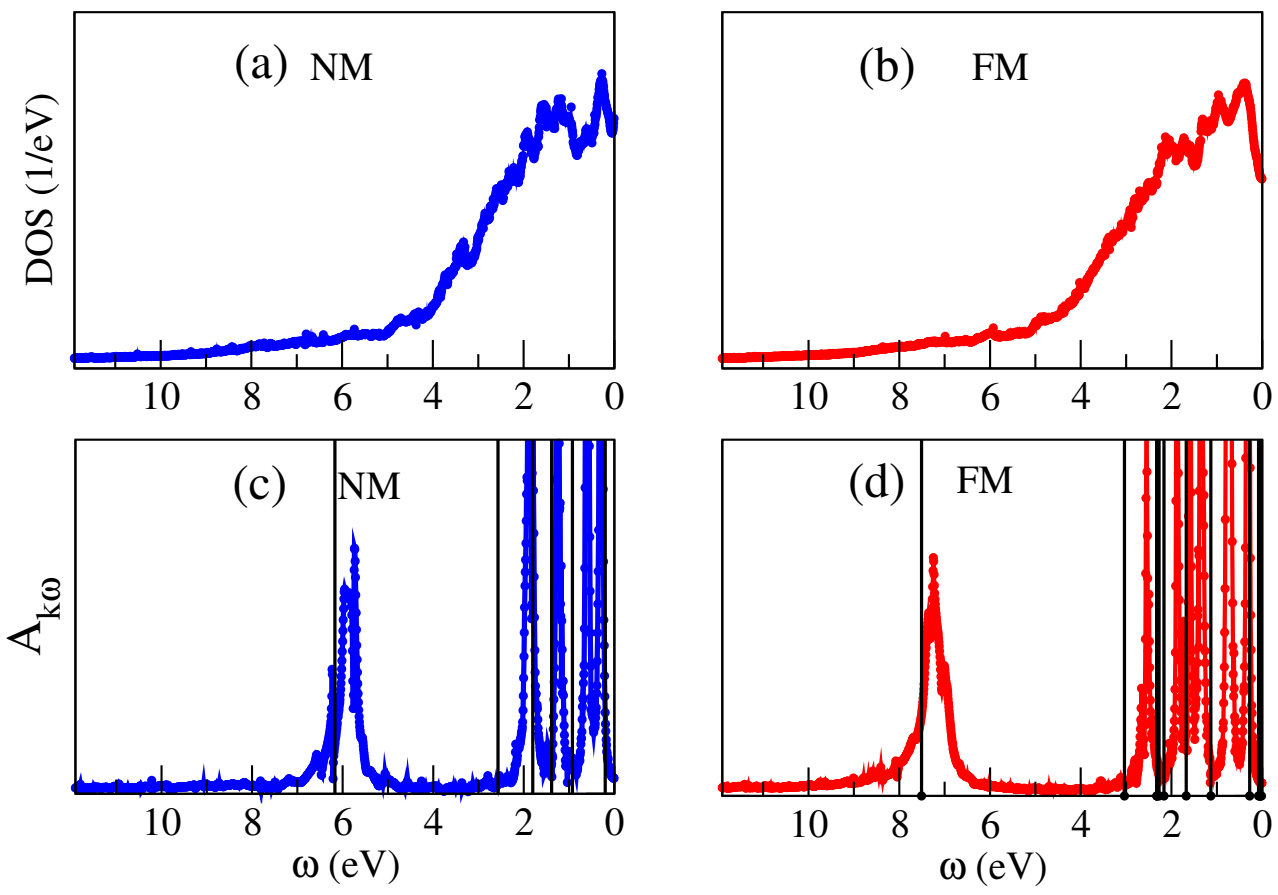

FIG. 11: (a) and (b) show total denisty of states plot of Ni in FM and NM phases respectively, using $G_0W_0$. (c) and (d) show the spectral function plot at a single k-point with coordinate (0.333, 0.166, 0.083) in FM and NM phases respectively, a representative of 6 $eV$ satellite feature obtained using $G_0W_0$.

## IV. CONCLUSION

In conclusion, we have revisited the electronic structure study of Ni to account for its observed electronic and magnetic properties. We have investigated its finite-temperature magnentic properties via a comparative DFT and DFT+DMFT study. We argue that the calculated Coulomb interaction parameters $U$ (5.78 $eV$) and $J$ (1.1 $eV$) using first principles approach along with rotationally invariant form of Coulomb paramterization in DFT+DMFT, explain both the electronic and magnetic properties of Ni to a remarkable extent. In comparison to DFT+DMFT, the DFT calculated finite temperature magnentism is found to well describe the experimental magnetization curve $M(T)$ of Ni with a theoretically proposed temperature scaling factor. We systematically show the failure of Stoner and spin fluctuation models of magnetism to explain its observed $M(T)$ curve.

Subsequently, significant contribution of Pauli-spin susceptibility along with Curie-Weiss suscpetibility to the total paramagnetic spin susceptibility has been observed. Our static mean-field finite-temperature method for studying magnetism in particular, finds importance due to consideration of temperature effects in self-consistent manner through the application of Fermi-dirac distribution function. The results establish the presence of comparative degrees of both the itinerant and local moment picture dictating the electronic structure properties in this system. Furthermore, the quasiparticle scattering rate for the Ni $3d$ states shows appreciable deviation from the conventional Fermi-liquid behavior across broad temperature range suggesting non-Fermi-liquid behavior. Our study finds presence of mixed valence electronic configuration ($3d^8$, $3d^9$ and $3d^7$) in the system. Interestingly, our DFT+DMFT calculated EES is able to account for the satellite feature $\sim$10 $eV$ observed in its resonant photoemission spectra along with the famous 6 $eV$ satellite feature. Our $G_0W_0$ results find the presence of plasmonic excitations contribution to the 6 $eV$ satellite feature along with the electronic correlation effects.

## REFERENCES

* Electronic mail: sbhardwajiitm369@gmail.com
† Current address: Department of Physics and Astronomy, West Virginia University, Morgantown, West Virginia 26506, USA
‡ Electronic mail: sudhir@iitmandi.ac.in